\documentclass[osajnl]{revtex4}  
\DeclareRobustCommand{\baselinestretch{2}}

\usepackage{graphicx}

\begin{document}

%
%
%





%
%
%

\title{Entanglement, Teleportation, And Single Photon Storage Using Two Level Atoms Inside An Optical Parametric Oscillator}


\author{P. R. Rice}%
 \email{ricepr@muohio.edu}
\affiliation{ Department of Physics, Miami University, Oxford,
Ohio 45056}%

\date{\today}

\begin{abstract}
I consider several interesting aspects of a new light source, a two-level atom, or N two-level atoms inside an Optical Parametric Oscillator. We find that in the weak driving limit, detection of a transmitted or fluorescent photon generates a highly entangled state of the atom and the cavity. This entanglement can be used with beamsplitters to create more complex quantum states and implement teleportation protocols. Also, one can store a single photon in the atoms, along the lines of recent slow and stopped light proposals and experiments.
\end{abstract}


\keywords{Photon Statistics, Optical Parametric Oscillator, Cavity
Quantum Electrodynamics, Coherence and Correlation Functions,
Squeezing }
  \maketitle
\section{INTRODUCTION}
For purposes of quantum computation and communication, it would be
useful to generate entangled states of quantum systems, and share
that entanglement \cite{book}. Generally dissipation destroys
quantum coherence, which is key to most proposed systems. A
measurement can project the system into an entangled state.
Cabrillo et. al. investigated a two atom system, driven
weakly\cite{Cabrillo}. The probability of either atom to be in the
excited state is then small, but not zero. If a spontaneous
emission event is detected without gaining information about which
atom emitted, the result is an entangled state of the two atoms. A
Bell measurement was used by Julsgaard et. al. to prepare two
Cesium cells in an entangled state\cite{Eugene}. Knill and
coworkers have proposed a quantum computing strategy that uses
only linear optics, single photon sources, and detectors, based on
this type of measurement induced entanglement \cite{Knill}.
Continuous measurement has been proposed to generate entanglement
\cite{Plenio}, which can then be used for teleportation
\cite{Bose}.  In the work of Duan et. al. \cite{Duan}, the output of two atomic cells that
were weakly pumped were coupled by a beamsplitter, with two
detectors behind it. The cells contain a large number of atoms
with a $\Lambda$ energy structure. Rarely the atoms in one of the
cells would make a Raman transition that results in the emission
of a photon. As the beamsplitter \lq\lq hides" knowledge of which
system emitted the photon, an entangled state of the two cells is
generated. That state is metastable, and to generate it one need
merely wait for a detector click. This system has been recently
implemented\cite{Lukin,HJK}.

There has also been interest in entanglement between two different types of systems, as in
the recent experiment of Blinov et. al. \cite{Blinov}. In their work,

   Duan and Kimble \cite{dk1} have recently proposed a similar scheme to
entangle many atoms. These schemes are all probabilistic, and one
must wait on a \lq\lq click" to prepare the system, but there is
no need for entanglement distillation and other complicated
procedures. Other schemes that utilize a measurement to implement
generation of entanglement and quantum computation include
Sorensen and Molmer \cite{Molmer}
 Marr et. al. have proposed an
adiabatic process of state preparation that is assisted by
dissipation\cite{Marr}. Zou et. al. have developed a scheme to
prepare a GHZ state using cavity decay \cite{GHZ}. Beige et. al.
have proposed using dissipation to maintain a system in a
decoherence free subspace \cite{Beige}.

Here I consider a similar scheme, with a
different optical system, that generates a wide variety of
entangled states. In this work I describe how detection of a photon from a weakly
driven cavity containing a nonlinear element and a two-level atom
results in an entangled state between the atom and cavity mode. By
then letting the system evolve we can generate a wide variety of
states. Then using two such systems coupled to detectors via a
beamsplitter this entanglement can be shared and used for teleportation. Also
we show that this system can be used to \lq\lq stop" light at the single photon level \cite{stoplight}. In section 2 we
describe the system under study and the generation and sharing of
entanglement. In section 3 we describe the use of this entanglement for teleportation and repeater applications, and section 4 discusses the light storage scheme at
the single-photon level, and I conclude in section 5.

\section{TWO LEVEL ATOM IN AN OPO}
We consider a single two-level atom inside an optical cavity,
which also contains a material with
 a $\chi^{(2)}$ nonlinearity. The atom and cavity are assumed to be resonant at $\omega$ and the
system is driven by light at $2\omega$. The system is shown in
Figure 1.
\begin{figure}[here]
   \begin{center}
   \begin{tabular}{c}
   \includegraphics[height=5cm]{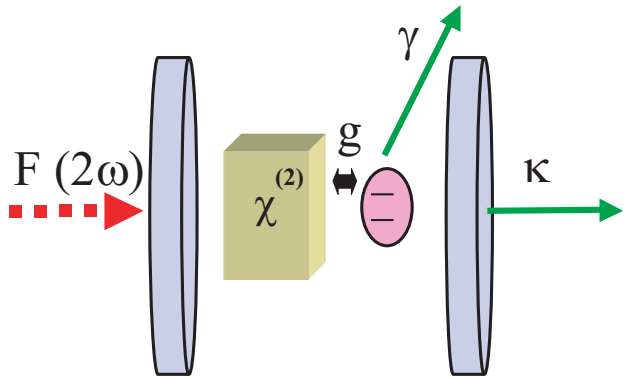}
   \end{tabular}
   \end{center}
   \caption[example]
{Two-level atom inside a driven optical parametric oscillator.
$F^2$ is the input photon flux at frequency $2\omega$, $g$ is the
atom-field coupling, $\gamma$ is the spontaneous emission rate out
the sides of the cavity, and $2\kappa$ is the rate of intracavity
intensity decay}
   \end{figure}

   The interaction of this driving field with the nonlinear
material produces light at the sub-harmonic $\omega$. This light
consists of correlated pairs of photons, or quadrature squeezed
light. In the limit of weak driving fields, these correlated pairs
are created in the cavity and eventually two photons leave the
cavity through the end mirror or as fluorescence out the side
before the next pair is generated. Hence we may view the system as
an atom-cavity system driven by the occasional pair of correlated
photons. This is in contrast to shining weakly squeezed light onto
the cavity, as it is not certain in that case that both entangled
photons get into the cavity.

In the language of squeezed light, we are interested in the limit
$N\rightarrow 0$. As $N$ is increased the effects we consider here
vanish. We wish to understand these effects in terms of photon
correlations rather than the usual effects of quadrature squeezed
light, where typically the largest nonclassical effects are seen
in the large $N$ limit.

We describe the system by a conditioned wave function,
 which evolves via a non-Hermitian Hamiltonian, and associated collapse processes.\cite{HJCbook} These are given by
\begin{eqnarray}
|\psi _c(t)\rangle &=&\sum\limits_{n=0}^\infty
{C_{g,n}(t)e^{-iE_{g,n}t}|g,n\rangle
+C_{e,n}(t)e^{-iE_{e,n}t}|e,n\rangle }\\ H&=& -i\kappa a^\dagger
a  -i{{\gamma}\over 2} \sigma_+\sigma_- +i\hbar F({a^{\dagger}}
^2-a^2)+i\hbar g(a^\dagger \sigma _--a\sigma _+)
\end{eqnarray}
where we also have collapse operators
\begin{eqnarray}
\cal{C}&=&\sqrt{\kappa} a\\
\cal{A}&=&\sqrt{{\gamma}\over2}\sigma_-.
\end{eqnarray}
representing cavity loss and spontaneous emission respectively.
Here, $g=\mu{({\omega}_0/\hbar{\epsilon}_0V)}^{1/2} $ is the usual
Jaynes-Cummings atom-field coupling in the rotating wave and
dipole approximations. The cavity-mode volume is $V$, and the
atomic dipole matrix element connecting ground and excited states
is $\mu$  . The effective two-photon driving field $F$  is
proportional to the intensity $I_{in}(2\omega_0)$ of a driving
field at  twice the resonant frequency of the atom (and resonant
cavity)  and the $\chi^{(2)}$  of the nonlinear crystal in the
cavity, as

\begin{equation}
F=-i\kappa _{in}\left( {{\cal{F} \over \pi }} \right)\sqrt
{{{\varepsilon _0V\;T} \over {\hbar \omega }}}e^{i\phi }\chi
^{(2)}I_{in}(2\omega )
\end{equation}

The cavity finesse is $\cal{F}$, $T$ and $\phi$ are the intensity
transmission coefficient and phase change at the input mirror. We
also have $\kappa_{in}=cT'/L$ as the cavity field loss rate
through the input mirror. The transmission $T'$ of the input
mirror is taken to be vanishingly small, with a large
$I_{in}(2\omega_0)$ so that $F$ is finite. Hence we effectively
consider a single ended cavity. Also $\gamma$ is the spontaneous
emission rate to all modes other than the privileged  cavity mode,
hereafter referred to as the vacuum modes. The field decay rate of
the cavity at the output mirror is $\kappa$. As we are working in
the weak driving field limit, we only consider states of the
system with up to 2 quanta, i.e.
\begin{equation}
|0-\rangle ,\;|0+\rangle ,\;|1-\rangle ,\;|1+\rangle ,\;|2-\rangle
 \label{eq:basisreal}
\end{equation}
Here, the first index corresponds to the excitation of the field
(n=number of quanta) and the second index denotes the number of
energy quanta in the atoms (+ for ground state, and - for excited
state).

We now construct an analytic solution using the quantum trajectory
method, and again look at weak driving fields. We need only keep the states with two or
less excitations (total in the cavity mode or internal energy) for
weak driving fields.

The equations for the relevant probability amplitudes are
\begin{eqnarray}
\dot{C}_{g,0}&=&-FC_{g,2}\nonumber\\
\dot{C}_{g,1}&=&gC_{e,0}-\kappa C_{g,1}\nonumber\\
\dot{C}_{e,0}&=&-gC_{g,1}-\gamma/2C_{e,0}\\
\dot{C}_{g,2}&=&g\sqrt{2}C_{e,1}+FC_{g,0}-2\kappa C_{g,2}\nonumber\\
\dot{C}_{e,1}&=&-\sqrt{2}gC_{g,2}-(\kappa
+\gamma/2)C_{e,1}\nonumber
\end{eqnarray}

We assume that the system starts in the ground state, and that
$C_{g,0}\sim 1$ for weak fields. After a collapse, the wave
function will evolve from the collapsed state back to the steady
state. The $0$- and $2$-photon amplitudes scale as $F$. As the
$1$-photon amplitudes are driven by decay from the $2$-photon
states, we assume that they also scale as $F$.   The solution to
these is (in unnormalized form)
\begin{eqnarray}
C_{g,1}(\tau)&=&exp\left(-\left({{\kappa}\over2}+{{\gamma}\over4}\right)\tau
\right)\Big[C_{g,1}(0)\cosh(\Omega \tau/2)  \nonumber
 \\&& +2{{\left(gC_{e,0}(0)-\left({{\kappa}\over2}-{{\gamma}\over4}\right)C_{g,1}(0)\right)}\over{\Omega}}\sinh(\Omega
\tau/2)\Big]\label{cg1}\\
C_{e,0}(\tau)&=&exp\left(-\left({{\kappa}\over2}+{{\gamma}\over4}\right)\tau
\right)\Big[ C_{e,0}(0)\cosh(\Omega \tau/2)  \nonumber
\\ && +2{{\left(\left({{\kappa}\over2}-{{\gamma}\over4}\right)C_{e,0}(0)-gC_{g,1}(0)\right)}\over{\Omega}}\sinh(\Omega
\tau/2) \Big]\label{ce0}
\end{eqnarray}
with
 \begin{equation}
 \Omega=\sqrt{{\left(\kappa-\gamma/2\right)}^2-4g^2}
 \end{equation}

In the steady state, one has (to order $F$)
\begin{eqnarray}
C_{g,0}^{SS}&=&1\nonumber\\
C_{g,1}^{SS}&=&0\nonumber\\
C_{e,0}^{SS}&=&0\\
C_{g,2}^{SS}&=&{F\over2}{{\kappa +\gamma/2}\over{g^2+\kappa(\kappa+\gamma/2)}}\nonumber\\
C_{e,1}^{SS}&=&{{-1}\over{\sqrt{2}}}{{gF}\over{g^2+\kappa(\kappa+\gamma/2)}}\nonumber
\end{eqnarray}
 The steady state photon number is given by
 \begin{equation}
\langle{\hat{a}}^{\dagger}\hat{a}\rangle=2{\mid
C_{g,2}^{SS}\mid}^2+{\mid C_{e,1}^{SS}\mid}^2\sim F^2
\end{equation}

 For an initial trigger
detection in the transmitted field, the appropriate collapsed
state is given by

\begin{eqnarray}
|\psi_{c}^T\rangle&\equiv &a|\psi \rangle_{SS}/|a|\psi\rangle_{SS}|\nonumber \\
&=&{{\sqrt{2}C_{g,2}^{SS}\mid g,1 \rangle+C_{e,1}^{SS}\mid e,0
\rangle}\over{\sqrt{2{\mid C_{g,2}^{SS}\mid}^2+{\mid
C_{e,1}^{SS}\mid}^2}}}\nonumber \\
&=&C_{g,1}^C\mid g,1 \rangle+C_{e,0}^C\mid e,0 \rangle
\end{eqnarray}
with
\begin{eqnarray}
C_{g,1}^C&=&{{\kappa+\gamma/2}\over{\sqrt{(\kappa+\gamma/2)^2+g^2}}}\\
C_{e,0}^C&=&{{-g}\over{\sqrt{(\kappa+\gamma/2)^2+g^2}}}
\end{eqnarray}

Note there is no population in the ground state. Upon detection of
a transmitted photon, as they are created in pairs, we find
ourselves certain in the knowledge that one quanta is in the
system, either in a cavity mode excitation (photon) or an internal
excitation of the atom.
For $g=\kappa$ and $\gamma=0.0$, we have a Bell state of the atom and field,
\begin{equation}
|\psi_c^T\rangle={1\over{\sqrt{2}}}\left(|0,e\rangle -|1,g\rangle\right)
\end{equation}
We can extend the type of state created by
detuning the atom and/or cavity by $\Phi$ or $\Delta$. The
probability amplitudes are then the same, with the replacement
\begin{eqnarray}
\kappa& \rightarrow & \kappa(1+i\Phi)\\
\gamma/2&\rightarrow & (\gamma/2)(1+i\Delta)
\end{eqnarray}
where $\Phi$ and $\Delta$ are the cavity and atom detunings scaled
by $\kappa$ and $\gamma/2$ respectively.

 As
shown in Fig. 2, after a detector click two such systems coupled by a beamsplitter yields
the state,
\begin{eqnarray}
|\psi_{c}^T\rangle&=&{1\over{\sqrt{2}}}\left[C_{g,1}^C(\mid
g,1;g,0\rangle +\mid g,0;g,1\rangle)\right. \nonumber
\\&&\left. +C_{e,0}^C(\mid e,0;g,0\rangle
+\mid g,0;e,0\rangle)\right]
\end{eqnarray}
, recalling that to order $F$ the state of the
uncollapsed system is $\mid 0,g\rangle$
\begin{figure}[here]
   \begin{center}
   \begin{tabular}{c}
   \includegraphics[height=5cm]{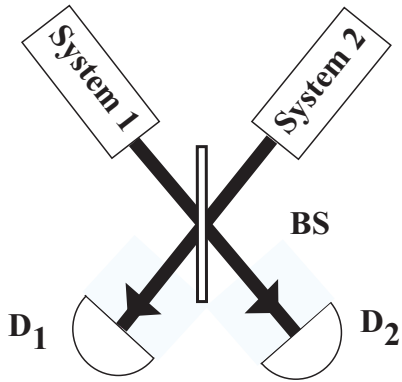}
   \end{tabular}
   \end{center}
   \caption[example]
{Two optical systems and two detectors coupled via a beamsplitter}
   \end{figure}
It is rare that both systems would emit a photon at (nearly) the
same time, as the rate of such emission is small in the weak field
limit, $2\kappa \langle a^{\dagger}a \rangle \Delta t \ll 1$. So
the first detection prepares the system in a highly entangled
state of two cavity modes and two atoms. One merely has to start
the system up, and wait for the first click, then the state is
prepared.

After this first detection, the state will evolve in time via Eqs.
(\ref{cg1}-\ref{ce0}).

Of particular interest is the case where $2\kappa=\gamma=g$, where
we have (after normalization)
\begin{eqnarray}
C^C_{g1}(\tau)&=&\left[
cos(g\tau)-sin(g\tau)\right]\\
C^C_{e0}(\tau)&=&\left[ cos(g\tau)+sin(g\tau)\right]
\end{eqnarray}
 This can  be used to generate different states, merely by choosing
 the delay time appropriate to the desired state. By detuning the
 cavity at a particular time, or by moving the atom to an antinode
 of the field using a moving lattice, one can \lq\lq freeze" this
 state.
 However, as opposed to the scheme of Duan et.
 al.\cite{Duan}
 where fewer states can be generated, this scheme does not generate metastable states.
 In our case the ground state contribution to
 this state is of order $F^2$ for times of the order of $1/\kappa$, and can still be neglected.
 However at some point, on the time scale of $1/\kappa$, a second
 photon will emerge from the cavity destroying the entanglement.
 A spontaneous emission event will do the same. As spontaneous emission events are harder to monitor, it will be advantageous in many situations to have $\gamma/\kappa\rightarrow 0$, so that the cavity decay mode dominates.

 In the limit of small $g$, we recover a regular, in which case the
 state of one system after a detection event is $|g,1\rangle$; when
 two systems are combined with a beamsplitter and two detectors as
 in Fig. 2, a detection event generates the state
 \begin{equation}
 |\Psi\rangle={1\over{\sqrt{2}}}(|01\rangle+|10\rangle)\label{2}
 \end{equation}
 with both atoms in the ground state (or indeed missing in the
 limit $g\rightarrow0$). Such entanglement could be generated
 simply by a pair of OPO's driven well below threshold. This way
 one could have two OPO's separated by some distance, pumped
 weakly, and coupled via a beam splitter and two detectors. When
 one detector fires, the two OPO's share an entangled state, which
 then can be used in the usual teleportation and dense coding
 schemes.
 This state will only live for a time on the order of $1/\kappa$,
 so it will result in probabilistic teleportation.

This scheme will work for $N$ atoms as well. In the weak field
limit, one additional state must be kept, that with two
excitations in the atoms and zero photons. Upon detection in
transmission, the resulting state $|\Psi\rangle_C$ will not
contain a contribution from that state as $a|0\rangle=0$. The
collapsed state will change in detail, most importantly by
the replacement of $g$ by $\sqrt{N}g$.

\section{Single Photon Storage}
In the case of a detection in fluorescence, the atom is in the ground state, and there is definitely one photon in the cavity mode. Also, if $\gamma \ll \kappa, g$, the resultant state upon detection in transmission is well approximated by $|1,g\rangle$ as well. This could be useful as a heralded photon state, but here we show that this is exactly the regime where one can store a single photon in the atom (or atoms), as in recent \lq\lq light storage" schemes \cite{stoplight}. These schemes rely on electromagnetically induced transparency (EIT). The EIT effect makes the three level atoms transparent to light on transition 1 in Fig. (3). As the light pulse begins to propagate in the medium, one turns the coupling laser off adiabatically, thereby transferring the coherent laser pulse  oscillation into a dipole oscillation between levels 2 and 3. Turning the coupling field $E$ back on slowly \lq\lq releases" the light pulse, with the phase preserved. An atom in a cavity, in the regime where there is one photon at most in the system is equivalent to a three-level $V$ atom, as in Fig. (3).\cite{CIT} For the three level atom, the EIT condition is $\gamma_1\gg E\gg \gamma_2$. Briefly this allows many Rabi oscillations between levels 2 and 3, while keeping them indistinguishable, setting the stage for quantum interference leading to EIT. One can of course understand this using coupled oscillators, which is why it is still observed relatively easily for many atoms and in solids.\cite{SHO}

\begin{figure}[here]
   \begin{center}
   \begin{tabular}{c}
   \includegraphics[height=5cm]{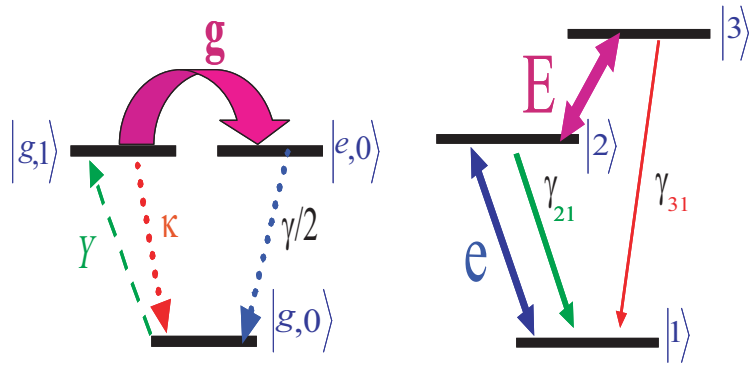}
   \end{tabular}
   \end{center}
   \caption[example]
{Equivalence of EIT and CIT systems}
   \end{figure}
The CIT condition is $\kappa \gg \sqrt{N}g \gg \gamma$. In this case the state of the system after detection of a photon in transmission is $|1,g\rangle$. One can then turn the CIT effect off, by adiabatically reducing the atom-field coupling $g$. This can perhaps be accomplished by holding the atoms in the OPO in a FORT or Optical Lattice, and then moving the minimum of the optical potential by moving the focus of the FORT or by chirping the lattice for a short period. This has the effect of taking the single photon Fock state and storing it as a dark state polariton of the atom-field system. The polariton can be released back into a cavity mode Fock state by moving the atom back into the field mode. This can be used to generate a photon on demand in the following way. Detect a photon in transmission in the CIT regime; adiabatically turn off $g$; turn $g$ back on at some later time. In this regime, the atom has a long lifetime relative to the cavity lifetime, and so one can have $\kappa$ relatively large to enhance the chance of a transmitted detection, and then use the longer atom lifetime to increase the single photon storage time.
 \section{CONCLUSION}
I have considered a system of a two-level atom (or atoms) in an optical cavity with a $\chi^{(2)}$ crystal inside, driven weakly at twice the atom-cavity resonant frequency. It would seem difficult to place a nonlinear crystal inside a high-Q cavity, but one should consider that the nonlinearity must be very weak, and one can use many atoms resulting in $g\rightarrow \sqrt{N}g$, and so the single photon coupling need not be the dominant rate. This system generates interesting entangled states upon detection of a transmitted photon. The generated state will
change with time, and eventually disappear, but is prepared with high fidelity in a probabilistic manner. Coupling such systems together as in previous schemes allows one to generate a larger variety of entantled states; this method can be used to entangle systems spatially separated, and can be used in teleportation and dense coding protocols. In one particular regime, a detection event prepares the system in a product state, albeit with a one-photon state for the cavity mode. If the atom(s) are slowly moved out of the cavity mode, perhaps with an optical lattice, one can implement \lq\lq light storage" at the single photon level. In this regime, the spontaneous emission rate is the smallest rate, and the inverse of that rate limits the storage time. The atom(s) can then be moved back into the field mode slowly, re-preparing the Fock state in the cavity mode. This scheme would be
difficult to implement, but it is hoped that it will spur
discussion of other schemes that will be more robust.

\end{document}